\documentclass[prd,a4paper,nofootinbib,superscriptaddress,preprintnumbers]{revtex4}
\usepackage[a4paper, hdivide={3cm,,3cm}, vdivide={3cm,,3cm}]{geometry}
\usepackage{amstext,amssymb,amsmath}
\usepackage[english]{babel}
\usepackage[ansinew]{inputenc}
\usepackage{units}
\usepackage{hyperref}
\usepackage[dvips]{color}

\begin{document}

\let \arrowvec \vec
\def \vec#1{{\boldsymbol{#1}}}
\def\ra{\rightarrow}
\newcommand{\matrixx}[1]{\begin{pmatrix} #1 \end{pmatrix}} 
\newcommand{\sm}{\mathrm{SM}}
\newcommand{\diag}{\mathrm{diag}}
\newcommand{\G}{G_\mathrm{max}}

\hypersetup{
    pdftitle={Vanishing Minors in the Neutrino Mass Matrix from Abelian Gauge Symmetries},
    pdfauthor={Takeshi Araki, Julian Heeck, Jisuke Kubo}
}

\title{Vanishing Minors in the Neutrino Mass Matrix\\ from Abelian Gauge Symmetries}

\author{Takeshi \surname{Araki}}
\email{araki@ihep.ac.cn}
\affiliation{Institute of High Energy Physics, Chinese Academy of Sciences, Beijing 100049, China}

\author{Julian \surname{Heeck}}
\email{julian.heeck@mpi-hd.mpg.de}
\affiliation{Max--Planck--Institut f\"ur Kernphysik, Saupfercheckweg 1, 69117 Heidelberg, Germany}
\affiliation{Institute for Theoretical Physics, Kanazawa University, Kanazawa 920-1192, Japan}

\author{Jisuke \surname{Kubo}}
\email{jik@hep.s.kanazawa-u.ac.jp}
\affiliation{Institute for Theoretical Physics, Kanazawa University, Kanazawa 920-1192, Japan}

\preprint{KANAZAWA-12-03}
\preprint{March 2012}


\begin{abstract}
Augmenting the Standard Model by three right-handed neutrinos allows for an anomaly-free gauge group extension $G_\mathrm{max} = U(1)_{B-L}\times U(1)_{L_e-L_\mu} \times U(1)_{L_\mu-L_\tau}$. While simple $U(1)$ subgroups of $G_\mathrm{max}$ have already been discussed in the context of approximate flavor symmetries, we show how two-zero textures in the right-handed neutrino Majorana mass matrix can be enforced by the flavor symmetry, which is spontaneously broken very economically by singlet scalars. These zeros lead to two vanishing minors in the low-energy neutrino mass matrix after the seesaw mechanism. This study may provide a new testing ground for a zero-texture approach: the different classes of two-zero textures with almost identical neutrino oscillation phenomenology can in principle be distinguished by their different $Z'$~interactions at colliders.
\end{abstract}

\maketitle


\section{Introduction}

In spite of many phenomenological successes, the Standard Model of particle physics~(SM) shows various problems that make an extension desireable.
One of the problems is the lack of a guidance principle governing the flavor structure of fermions: the SM gauge symmetries are incapable to explain the observed mixing patterns and mass spectra of the quarks and leptons. In order to remedy this incompleteness, global family symmetries are often introduced in the literature, usually discrete instead of continuous. Another popular ansatz is to extend the gauge sector to include a new gauge quantum number which distinguishes the three families of fermions.
Family non-universally interacting new $U(1)$ gauge symmetries are extensively studied in Refs.~\cite{non-uni} with a focus on flavor changing neutral current (FCNC) processes mediated by the new gauge boson.
In this letter, we will instead focus on the flavor structure of neutrinos in the framework of the type-I seesaw mechanism~\cite{type1}.
We will first classify possible anomaly-free $U(1)$ gauge symmetries and then show that such $U(1)$ gauge symmetries result in diagonal Dirac mass matrices and texture zeros in the right-handed Majorana neutrino mass matrix, which in turn give vanishing minors in the low-energy neutrino mass matrix~\cite{0minr,0minr2, moretexturezeros, 0texr}.
We find that all of the phenomenologically consistent seven patterns of two vanishing minors can naturally be obtained in the presence of one or two SM gauge singlet scalars, while the occurring gauge boson $Z'$ provides a handle to distinguish the different patterns outside of the neutrino sector.

The symmetry realization of texture zeros is well studied in Refs.~\cite{Zn}, see also Refs.~\cite{0minr,0minr2, moretexturezeros, 0texr}, by means of discrete $\mathbb{Z}_N$ symmetries.
However, it is not completely clear that discrete global symmetries would survive gravitational effects such as wormholes~\cite{gravity}, and moreover the spontaneous breaking of discrete symmetries may suffer from the domain wall problem~\cite{dwall}.
In this sense, it might be more convincing to adopt gauge symmetries instead of discrete ones.
Furthermore, the new $U(1)$ gauge bosons can be expected to have some impacts on the LHC phenomenology and therefore provide better testability. 

The remainder of this letter is organized as follows. We discuss the extension of the SM by the abelian gauge group $\G=U(1)_{B-L}\times U(1)_{L_e-L_\mu} \times U(1)_{L_\mu-L_\tau}$ in Sec.~\ref{sec:maximalabeliangroup}. Taking simple $U(1)$ subgroups of $\G$, we show how to construct texture zeros in the Majorana mass matrix of the right-handed neutrinos in Sec.~\ref{sec:texturezeros}, which lead to vanishing minors in the low-energy neutrino mass matrix. Our findings are concluded and summarized in Sec.~\ref{sec:conclusion}. The notation for texture zeros and minors is given in App.~\ref{sec:patterns}. App.~\ref{sec:lists} lists all flavor symmetries that result in valid two-zero textures with the addition of two or three SM singlet scalars.

\section{Maximal Abelian Gauge Group}
\label{sec:maximalabeliangroup}

We extend the SM by introducing a new $U(1)$ gauge symmetry, whose charge is denoted by $Y'$, and three right-handed Majorana neutrinos $N_R$ to implement the type-I seesaw mechanism.
The $Y^\prime$ assignments for the quarks are assumed to be family universal to avoid dangerous FCNCs in the quark sector, while those for the leptons are family non-universal, i.e.~$Y^\prime(q_L)=-Y^\prime(u_R^c)=-Y^\prime(d_R^c)$ for all families while $Y^\prime(\ell_{Li})=-Y^\prime(e_{Ri}^c)=-Y^\prime(N_{Ri}^c)$ and $Y^\prime(\ell_{Li}) \neq Y^\prime(\ell_{Lj})$ for each family.
We keep the SM Higgs neutral under the new $U(1)$ gauge symmetry.

The cubic and gravitational anomalies of the new $U(1)$ gauge symmetry vanish automatically due to the vectorlike structure of the assigned fermion charges in the presence of the three right-handed neutrinos. This leaves the possible anomalies with the SM gauge group, the representations of which are given in Tab.~\ref{tab:hypercharges}, leading to the sole condition 
\begin{align}
 9\,  Y' ( q_L )  +Y' (\ell_{L1}) +Y' (\ell_{L2})+Y' (\ell_{L3})= 0\,.
 \label{eq:anomalyconstraint}
\end{align}
If $Y' ( q_L )$ is non-zero, this condition leads to a $B-\sum_\alpha x_\alpha L_\alpha$ gauge symmetry with the constraint $\sum_\alpha x_\alpha = 3$~\cite{B-stuff}, where $B$ and $L_{\alpha=e,\mu,\tau}$ denote baryon and lepton-flavor numbers, respectively. The case of family universal charges gives the well-known $B - L$ gauge symmetry, $L$ being the lepton number $L= L_e + L_\mu + L_\tau$.
On the other hand, $Y' ( q_L ) = 0$ leads to a purely leptonic symmetry $\sum_\alpha y_\alpha L_\alpha$ with $\sum_\alpha y_\alpha =0$, which has been discussed briefly in Ref.~\cite{Davoudiasl:2011sz}.
As a special case, $y_e=0$, $y_\mu=0$ or $y_\tau =0$ correspond to a gauged lepton-number difference $L_\alpha-L_\beta$. These are actually anomaly-free within the SM~\cite{zero}, i.e.~without right-handed neutrinos, and have been discussed extensively in the literature~\cite{zero,Lmu-Ltau}.

\begin{table}[t]
\renewcommand{\baselinestretch}{1.6}\normalsize 
\centering
\begin{tabular}[t]{|l|l|l|}
\hline                               
     $\ell_{Li} = (\nu, \, e_L)^T_i \sim \left(\vec{1},\vec2,-\frac{1}{2}\right)$ & $e_{Ri}^c \sim \left(\vec1,\vec1,+1\right)$ & $N_{Ri}^c\sim \left(\vec1,\vec1,0\right)$\\\hline
     $q_{Li} = (u_L, \, d_L)^T_i \sim \left(\vec3,\vec2,+\frac{1}{6}\right)$ & $u_{Ri}^c \sim \left({\overline{\vec3}},\vec1,-\frac{4}{6}\right)$ & $d_{Ri}^c \sim \left(\overline{\vec3},\vec1,+\frac{2}{6}\right)$ \\
\hline 
\end{tabular}
\renewcommand{\baselinestretch}{1.0}\normalsize 
\caption{\label{tab:hypercharges}
$SU(3)_C \times SU(2)_L \times U(1)_Y \equiv G_\sm$ representations of left-handed fermions,
where $i=1 \dots 3$ denotes the family index. Electric charge is defined as $Q=T_3 +Y$.}
\end{table}

While the extension of the SM gauge group by a single new $U(1)$ gauge symmetry satisfying Eq.~\eqref{eq:anomalyconstraint} has been discussed before, it has to our knowledge never been emphasized that every such $U(1)$ can be viewed as a subgroup of $U(1)_{B-L} \times U(1)_{L_\alpha - L_\beta} \times U(1)_{L_\beta - L_\gamma}$, which is also free of anomalies. Since each factor on its own satisfies Eq.~\eqref{eq:anomalyconstraint}, only the cross terms could be problematic. We show the purely leptonic part
\begin{align}
 \sum &(L_\alpha - L_\beta)^2 (L_\beta - L_\gamma) = \sum_{\beta\text{ fermions}} L_\beta^3 = 2\times (+1)^3 + (-1)^3 + (-1)^3 = 0\,,\\
 \sum &(L_\alpha - L_\beta) (L_\beta - L_\gamma)^2 =- \sum_{\beta\text{ fermions}} L_\beta^3 = 0\,,\\
 \sum &(L_\alpha - L_\beta) (L_\beta - L_\gamma) Y =- \sum_{\beta\text{ fermions}} L_\beta^2 Y =-\left[ 2\times (+1)^2 \left(-\frac{1}{2}\right) + (-1)^2 (+1)\right] = 0\,,
\end{align}
and the anomalies involving $B-L$:
\begin{align}
 \sum &(B-L) (L_\alpha - L_\beta)^2 = -2\sum_{\beta\text{ fermions}} L_\beta= -2\left[ 2\times (+1) + (-1) + (-1)\right] = 0\,,\\
 \sum &(B-L) (L_\alpha - L_\beta) (L_\beta - L_\gamma) = \sum_{\beta\text{ fermions}} L_\beta^3= 0\,,\\
 \sum &(B-L) (L_\alpha - L_\beta) Y = \sum_{\alpha\text{ fermions}} (B-L) Y -\sum_{\beta\text{ fermions}} (B-L) Y= 0\,,\\
 \sum &(B-L)^2 (L_\alpha - L_\beta) = \sum_{\alpha\text{ fermions}} (B-L)^2 -\sum_{\beta\text{ fermions}} (B-L)^2= 0\,,
\end{align}
where the last two relations follow from the universality of $Y$ and $B-L$~\cite{kineticmixing}.
This means that the maximal anomaly-free abelian gauge group of the SM+$3\,N_R$ is given by
\begin{align}
 \G \equiv U(1)_{B-L} \times U(1)_{L_e - L_\mu} \times U(1)_{L_\mu - L_\tau}\,,
\label{eq:maximalsymmetrygroup}
\end{align}
where we chose a specific direction in group space. Let us make two comments about the obtained result.
\begin{itemize}
\item 
First note that even though we can formally consider
\begin{align}
\begin{split}
 G_\sm &\times U(1)_{B-x_e L_e - x_\mu L_\mu - x_\tau L_\tau}\times U(1)_{y_e L_e +y_\mu L_\mu +y_\tau L_\tau}  \\
&\times U(1)_{L_e - L_\mu} \times U(1)_{L_\mu - L_\tau}\times U(1)_{L_\tau- L_e}
\end{split}
\end{align}
and show that it is anomaly-free for $\sum x_\alpha = 3$ and $\sum y_\alpha = 0$, the decompositions
\begin{align}
  & L_e-L_\tau = (L_e - L_\mu) + (L_\mu - L_\tau) ,\\
& y_e L_e + y_\mu L_\mu - (y_e + y_\mu) L_\tau = y_e(L_e-L_\mu)+(y_e + y_\mu) (L_\mu-L_\tau) ,\\
& B-x_e L_e - x_\mu L_\mu - (3-x_e - x_\mu) L_\tau = (B-L) + (1-x_e) (L_e - L_\mu) \\
& \qquad\qquad\qquad\qquad\qquad\qquad\qquad\qquad\quad  + (2-x_e-x_\mu) (L_\mu - L_\tau) ,\nonumber
\end{align}
show that the generators of the five new abelian groups are not independent and only two of the lepton-number differences can be gauged.\footnote{Stated in another way: One of the gauge bosons of $U(1)_{L_e - L_\mu} \times U(1)_{L_\mu - L_\tau}\times U(1)_{L_\tau- L_e}$ can be rotated away, i.e.~made non-interacting, so it suffices to consider $U(1)_{L_e - L_\mu} \times U(1)_{L_\mu - L_\tau}$ (with kinetic mixing). The same argument holds for the other linear combinations.}
\item
Second, there is a more elegant way to derive the vanishing anomalies. By taking another basis for the flavor-dependent part of $\G$ (acting on flavor space), namely
\begin{align}
L_e - L_\mu = \diag (1, -1, 0) , && (L_e - L_\mu) + 2\,(L_\mu -L_\tau) = \diag (1,1,-2) ,
\end{align}
we see that these two generators form the Cartan subalgebra of a rank-2 $SU(3)_\ell$. Putting the leptons in the representations
\begin{align}
\ell_{Li} \sim \vec{3}_\ell\,, && e_{Ri}^c \sim \overline{\vec{3}}_\ell\,, && N_{Ri}^c \sim \overline{\vec{3}}_\ell\,,
\end{align}
immediately shows that they form a real reducible representation of $SU(3)_\ell$, so the anomaly $SU(3)_\ell$--$SU(3)_\ell$--$SU(3)_\ell$ vanishes, in direct analogy to the quarks in $SU(3)_C$. Anomalies with other nonabelian group factors vanish trivially, so the only possible anomalies are
\begin{align}
\begin{split}
 &SU(3)_\ell\text{--} SU(3)_\ell \text{--} U(1)_Y: \quad \sum_{\vec{3}_\ell} Y = 3\times \left(2\, Y(L_e) + Y (e_R^c)\right) = 0\,,\\
 &SU(3)_\ell\text{--} SU(3)_\ell \text{--} U(1)_{B-L}: \quad \sum_{\vec{3}_\ell} (B-L) = 3\times \left(2\, (-1) + (+1) + (+1)\right) = 0\,,
\end{split}
\end{align}
which means $G_\sm \times U(1)_{B-L} \times SU(3)_\ell$ is anomaly-free. 
Since the $SU(3)_\ell$ is badly broken by the Yukawa couplings in the charged lepton sector, we will not use it in the following.
\end{itemize}

To partially summarize: we have shown that with three right-handed neutrinos, the group
\begin{align}
 G_\sm \times \G = SU(3)_C \times SU(2)_L \times U(1)_Y \times U(1)_{B-L} \times U(1)_{L_e - L_\mu} \times U(1)_{L_\mu - L_\tau}\,,
\end{align}
is free of anomalies. Every subgroup of $\G$ is of course automatically anomaly-free, i.e.~satisfies Eq.~\eqref{eq:anomalyconstraint}, and many of them have already been discussed in the literature. 
The exact symmetry $\G$ forbids the Majorana mass terms of the right-handed neutrinos, so they have to be induced by breaking the group with SM singlet scalars similar to the usual $B-L$ models.\footnote{To illustrate this point: the coupling $S\, \overline{N}_1^c N_1$ is allowed if the complex scalar $S$ carries the charge $(+2, -2,0)$ under $U(1)_{B-L} \times U(1)_{L_e - L_\mu} \times U(1)_{L_\mu - L_\tau}$. For $S\, \overline{N}_2^c N_3$ the charge assignment is $(+2, +1,0)$. In both cases the $L_\mu-L_\tau$ symmetry remains unbroken after $S$ acquires a vacuum expectation value.} A discussion of the full breakdown $\G\rightarrow$~nothing---and its connection to neutrino mass and mixing---lies outside the realm of this letter, as it involves many parameters and new scalars. Instead, we focus on an effective model of a possible last step of the breakdown $\G\ra U(1)'$, i.e.~we consider only a $U(1)$ subgroup of $\G$. This has the advantage of a simple scalar sector with tree-level couplings to the right-handed neutrinos.

The question then arises which subgroup of $\G$ should be chosen. As far as approximate flavor symmetries in the neutrino mass matrix go, three interesting cases have been identified already in Ref.~\cite{Choubey:2004hn}, namely $L_e$ symmetry for normal mass ordering, $L_e-L_\mu-L_\tau$ for inverted mass ordering and $L_\mu-L_\tau$ for quasi-degenerate neutrinos. The first two can be promoted to a local symmetry with the help of the baryon number, i.e.~by considering $U(1)_{B-3 L_e}$ and $U(1)_{B+ 3 (L_e - L_\mu -L_\tau)}$~\cite{soon}, while the last one has long been used before~\cite{Lmu-Ltau}.
The motivation behind these symmetries is the possible approximate flavor symmetry in the neutrino mass matrix, i.e.~the symmetry violating entries in $\mathcal{M}_\nu$ can be significantly smaller than the symmetry preserving entries. While this can provide a motivation for neutrino hierarchies and the rough pattern of the mixing angles, the spontaneously broken symmetries usually do not lead to testable relations among the neutrino mixing parameters.

Here we will take a different approach and discuss flavor symmetries $U(1) \subset \G$ that generate texture zeros or vanishing minors in the neutrino mass matrices after breaking the gauge symmetry spontaneously by only a few scalars. Two independent zeros (or vanishing minors) in the active neutrino mass matrix imply four constraints on the nine low-energy parameters $(m_1,m_2,m_3)$, $(\theta_{23},\theta_{12},\theta_{13})$ and $(\delta,\alpha,\beta)$ (CP violating phases), making them in principle distinguishable with future data. Our approach not only provides new testing ground for flavor symmetries, but also allows to check for the flavor symmetry behind the texture zeros at the LHC.

\section{Two Vanishing Minors in the Neutrino Mass Matrix}
\label{sec:texturezeros}

After integrating out the heavy right-handed neutrinos, we obtain the low-energy neutrino mass matrix
\begin{align}
\mathcal{M}_\nu \simeq -m_D \mathcal{M}_R^{-1} m_D^T \,,
\end{align}
where $m_D$ and $\mathcal{M}_R$ are the Dirac and right-handed Majorana neutrino mass matrices, respectively.
Because we are considering family non-universal $U(1)$ gauge symmetries, $m_D$ and the charged lepton mass matrix are forced to be diagonal,\footnote{Note that our symmetries do not constrain the values of Yukawa couplings. In particular, we can not explain the hierarchy of the charged lepton masses in this framework, but have to put in the right Yukawa couplings by hand. An extension of our model by a Frogatt-Nielsen type mechanism~\cite{Froggatt:1978nt} to explain the hierarchy may be possible, but goes beyond the scope of this letter.} i.e.~$m_D = \diag (a,~b,~c)$, while usually several elements of $\mathcal{M}_R$ are forbidden by the flavor symmetry.
Some elements of $\mathcal{M}_R$ may be induced by introducing SM gauge singlet scalars with suitable $U(1)$ charges and non-zero vacuum expectation values (VEVs), but the other elements may remain vanishing.
Thus, we can discuss texture zeros in $\mathcal{M}_R$, which give rise to vanishing minors in $\mathcal{M}_\nu$.
As far as looking for vanishing minors in $\mathcal{M}_\nu$ is concerned, the values of $a,~b,~c$ in $m_D$ are irrelevant to the matter, so we can simply set $a=b=c=1$. We will comment on this choice at the end of this section.

The analysis of texture zeros in $\mathcal{M}_\nu$ has been recently performed in Ref.~\cite{0texr}, with the result that seven patterns of $\mathcal{M}_\nu$ with two independent zeros, that is  $\vec{A}_1^\nu$, $\vec{A}_2^\nu$, $\vec{B}_1^\nu$, $\vec{B}_2^\nu$,  $\vec{B}_3^\nu$, $\vec{B}_4^\nu$ and $\vec{C}^\nu$ (see App.~\ref{sec:patterns} for the notation), are consistent with the latest global fit of neutrino oscillation data at the $3\sigma$ level.
Of the seven patterns, $\vec{A}_2^\nu$, $\vec{A}_1^\nu$, $\vec{B}_4^\nu$ and $\vec{B}_3^\nu$ translate into the following two-zero textures in $\mathcal{M}_R$ (or $\mathcal{M}_\nu^{-1}$)
\begin{align}
 \vec{D}_1^R:\ \matrixx{\times & \times & \times \\ \cdot & 0 & 0 \\ \cdot & \cdot & \times}  , &&
 \vec{D}_2^R:\ \matrixx{\times & \times & \times \\ \cdot & \times & 0 \\ \cdot & \cdot & 0}  , &&
 \vec{B}_3^R:\ \matrixx{\times & 0 & \times \\ \cdot & 0 & \times \\ \cdot & \cdot & \times}  , &&
 \vec{B}_4^R:\ \matrixx{\times & \times & 0 \\ \cdot & \times & \times \\ \cdot & \cdot & 0}  ,
 \label{eq:texturezero}
\end{align}
respectively, while the other patterns $\vec{B}_1^\nu$, $\vec{B}_2^\nu$ and $\vec{C}^\nu$ do not lead to texture zeros in $\mathcal{M}_R$. Correspondingly, there might be additional allowed zeros in $\mathcal{M}_R$ that do not give zeros in $\mathcal{M}_\nu$ and are therefore invisible in the analysis of Ref.~\cite{0texr}. Using the current values for the mixing angles and mass-squared differences from Ref.~\cite{globalfit}, we checked that the following three patterns of $\mathcal{M}_R$
\begin{align}
 \vec{B}_1^R:\ \matrixx{\times & \times & 0 \\ \cdot & 0 & \times \\ \cdot & \cdot & \times}  , &&
 \vec{B}_2^R:\ \matrixx{\times & 0 & \times \\ \cdot & \times & \times \\ \cdot & \cdot & 0}  , &&
 \vec{C}^R:\ \matrixx{\times & \times & \times \\ \cdot & 0 & \times \\ \cdot & \cdot & 0}  ,
 \label{eq:texturezero2}
\end{align}
are also allowed at the $3\sigma$ level.
Consequently, we have seven allowed two-zero textures in $\mathcal{M}_R$ in analogy to $\mathcal{M}_\nu$.
For convenience we list the allowed two-zero textures in $\mathcal{M}_\nu$ and $\mathcal{M}_R$ in terms of the notation defined in App.~\ref{sec:patterns}:
\begin{align}
\begin{split}
\mathcal{M}_\nu &: \qquad \vec{A}_1^\nu\,,\vec{A}_2^\nu\,,\vec{B}_1^\nu\,,\vec{B}_2^\nu\,,\vec{B}_3^\nu\,,\vec{B}_4^\nu\,,\vec{C}^\nu\,,\\
\mathcal{M}_R &: \qquad \vec{D}_1^R\,,\vec{D}_2^R\,,\vec{B}_1^R\,,\vec{B}_2^R\,,\vec{B}_3^R\,,\vec{B}_4^R\,,\vec{C}^R\,.
\end{split}
\label{eq:allowedtexturezerosandminors}
\end{align}
Notice that the same conclusion for $\mathcal{M}_R$ was reached in Refs.~\cite{0minr} with old data, and $\vec{B}_1^R$ and $\vec{B}_2^R$ were closely discussed recently in Ref.~\cite{0minr2}.
Using the recent Daya Bay result~\cite{dayabay} for the reactor angle $\sin^2 2\theta_{13} =  0.092 \pm 0.017$ we can show that all the above patterns in $\mathcal{M}_R$ (Eqs.~\eqref{eq:texturezero} and~\eqref{eq:texturezero2}) are actually even valid at $1\sigma$.

The patterns $\vec{B}_i^R$ admit normal as well as inverted hierarchy solutions, while $\vec{D}_i^R$ and $\vec{C}^R$ require normal ordering. To illustrate how well the different textures perform, we filled the non-vanishing entries in $\mathcal{M}_R$ with random complex numbers of magnitude $\leq 1$ and checked if the resulting neutrino mass matrix has parameters $\theta_{ij}$, $\Delta m_{21}^2/\Delta m_{31}^2$ in the allowed $3\sigma$ range (using again the Daya Bay result for $\theta_{13}$). From the patterns $\vec{D}_i^R$, $\mathcal{O}( 10^6)$ out of $10^9$ random matrices were compatible with data, $\vec{C}^R$ gave $\mathcal{O}(10^4)$ valid matrices and the $\vec{B}_i$ patterns $\mathcal{O}(10^2)$.\footnote{The exact numbers $(\# \mathrm{NH},\# \mathrm{IH})$ of valid matrices for $10^9$ random tries were: $(2.9\times 10^6,0)$ for $\vec{D}_1^R$, $(2.8\times 10^6,0)$ for $\vec{D}_2^R$, $(7961,0)$ for $\vec{C}^R$, $(950,54)$ for $\vec{B}_1^R$, $(335,78)$ for $\vec{B}_2^R$, $(543,50)$ for $\vec{B}_3^R$ and $(215,80)$ for $\vec{B}_4^R$.}
A more detailed analysis of fine-tuning in $\mathcal{M}_\nu$ texture zeros was recently performed in Ref.~\cite{finetuning}, where the least fine-tuned patterns were identified as $\vec{A}_i^\nu$ (which is our $\vec{D}^R_j$). Since the $\mathcal{M}_R$ textures of Eq.~\eqref{eq:texturezero2} do not lead to texture zeros in $\mathcal{M}_\nu$, they were not considered in the analysis of Ref.~\cite{finetuning}. However, the counting of valid random matrices suggests a similar conclusion, i.e.~the patterns $\vec{B}_i^R$ and $\vec{C}^R$ can be considered less natural than $\vec{D}_i^R$, at least for normal hierarchy. Should the mass ordering of neutrinos be inverted, we would just have the $\vec{B}_i^R$ textures, with similar performance.

\subsection{Realization via Continuous Flavor Symmetries}

We will not discuss the implications of the texture zeros on the neutrino mixing parameters any further, see e.g.~Refs.~\cite{0minr} for dedicated detailed analyses. In the following we will rather show that all of the allowed patterns for $\mathcal{M}_R$ can be derived by family non-universal $U(1)$ gauge symmetries---discussed in Sec.~\ref{sec:maximalabeliangroup}---with at most two new SM singlet scalars.
To make the connection between texture zeros and symmetries, we list the charge-matrices of $Y'(\overline{N}^c_i N_j)$ for the two cases $Y'(q_L)=0$ and $Y'(q_L)\neq 0$, i.e.~$Y'=y_e L_e + y_\mu L_\mu -(y_e + y_\mu) L_\tau$ and $Y' =B-x_e L_e - x_\mu L_\mu - (3-x_e-x_\mu) L_\tau$:\footnote{Due to the insignificant overall normalization of $U(1)$ generators it is sufficient to consider these two two-parameter subgroups of $\G$. $Y'=y_e L_e + y_\mu L_\mu -(y_e + y_\mu) L_\tau$ could be similarly split into $y_e = 0$ ($Y' = L_\mu-L_\tau$) and $y_e\neq 0$ ($Y' = L_e + y_\mu L_\mu - (1+y_\mu) L_\tau$), which however barely simplifies matters.}
\begin{align}
\matrixx{2 y_e & y_e + y_\mu & -y_\mu \\ y_e + y_\mu & 2 y_\mu & -y_e\\ -y_\mu & -y_e & -2 (y_e+y_\mu)} , &&
\matrixx{-2 x_e & -x_e -x_\mu & x_\mu-3\\ -x_e - x_\mu & -2 x_\mu & x_e -3\\ x_\mu -3 & x_e -3 & 2 x_e + 2 x_\mu -6} .
\end{align}
We note that the family non-universality requires $y_e \neq y_\mu$, $y_e \neq -2y_\mu$, $y_\mu \neq -2y_e$, $x_e \neq x_\mu$, $x_e \neq 3-2x_\mu$ and $x_\mu \neq 3-2x_e$.
Imposing an exact $L_\alpha - L_\beta$ symmetry, for instance, results in $4$ zeros and two independent symmetry conserving entries (with scale $M_{L_\alpha-L_\beta}$). A scalar with $L_\alpha - L_\beta$ charge $ \pm 1$ or $\pm 2$ will fill two of those zeros after acquiring a VEV. Matching this to Eq.~\eqref{eq:texturezero} and Eq.~\eqref{eq:texturezero2} shows that only the $L_\mu - L_\tau$ symmetry with a scalar $S$ whose charge is $\pm 1$ can lead to a valid pattern ($\vec{C}^R$):
\begin{align}
 \mathcal{M}_R = M_{L_\mu-L_\tau} \matrixx{\times & 0 & 0 \\ \cdot & 0 & \times \\ \cdot & \cdot & 0} + \langle S \rangle \matrixx{0 & \times & \times \\ \cdot & 0 & 0 \\ \cdot & \cdot & 0} \sim  \matrixx{\times & \times & \times \\ \cdot & 0 & \times \\ \cdot & \cdot & 0}.
\end{align}
The remaining zeros in this case will be filled by effective operators $S^2\, \overline{N}_i^c N_j/\Lambda$, suppressed by a new physics scale $\Lambda$. In order for us to talk about texture ``zeros'', we need $\Lambda \gg M_{L_\mu-L_\tau},\langle S\rangle$. Furthermore, the charged lepton mass matrix will also receive off-diagonal elements suppressed by~$\Lambda^n$, which introduces a contribution $U_\ell$ to the lepton mixing matrix $U_\mathrm{PMNS} = U_\ell^\dagger U_\nu$. Correspondingly, the predictivity of the texture zero approach goes down the drain if we allow for a low $\Lambda$, but the perturbations could of course be used to alleviate any tension between the predicted and observed values. Since all the $U(1)$ models we employ in this paper are anomaly-free, our models are renormalizable and thus can be valid up to the Planck scale (assuming this is where quantum gravity takes over). In the following, we will therefore always assume these higher dimensional operators to be sufficiently suppressed.

Back to the possible flavor symmetries that give two vanishing minors in $\mathcal{M}_\nu$.
In the case of $y_e\neq 0$, $y_\mu \neq 0$ and $y_e \neq -y_\mu$, we need at least three SM singlet scalars in order to construct the allowed patterns of two-zero textures. This is of course not particularly exciting, so we only display them in App.~\ref{sec:lists}.

Going to the $B-x_e L_e - x_\mu L_\mu - (3-x_e-x_\mu) L_\tau$ symmetry allows for a lot more patterns; there are many assignments for $x_e$ and $x_\mu$ that give one or even no zeros and can therefore easily produce consistent phenomenology (see Ref.~\cite{one_vanishing_minor} for implications of just one vanishing minor in $\mathcal{M}_\nu$). There are also assignments that lead to valid two-zero textures with just one scalar, a complete list is shown in Tab.~\ref{tab:symmetryzeros}. We see that only the patterns $\vec{D}_1^R$, $\vec{D}_2^R$, $\vec{B}_3^R$, $\vec{B}_4^R$ and $\vec{C}^R$ can be obtained in this highly economic way.
The charge assignments are summarized in Tab.~\ref{tab:symmetryzeros} with the lower bounds on the breaking scale of the $U(1)$, $|M_{Z'}/g' | = |Y'(S) \,v_S| =\sqrt{2}\, |Y'(S) \,\langle S\rangle|$, as determined by the anomalous magnetic moment of the muon~\cite{Lmu-Ltau} or LEP-II measurements~\cite{LEP-2bounds}. 

If we extend the scalar sector by two SM singlet scalars instead of just one, we can construct the remaining two valid patterns of $\mathcal{M}_R$ listed in Eq.~\eqref{eq:texturezero2}, by using for example $B- L_e -5 L_\mu + 3 L_\tau$ for $\vec{B}_2^R$ and $B- L_e+3 L_\mu -5 L_\tau$ for $\vec{B}_1^R$, respectively. In both cases we need scalars with charge $|Y' (S_1)|=2$ and $|Y' (S_2)|=10$. Since there is no unique symmetry behind the patterns $\vec{B}_{1,2}^R$, we will not discuss them any further. 
App.~\ref{sec:lists} provides a complete list of the $B- x_e L_e - x_\mu L_\mu - (3-x_e - x_\mu) L_\tau$ charge assignments that yield the allowed two-zero textures in $\mathcal{M}_R$ with two scalars. Some of the solutions do not allow for flavor symmetric mass terms, which means there are only the two breaking scales $\langle S_1\rangle$ and $\langle S_2\rangle$ that determine $\mathcal{M}_R$.

\begin{table}[t]
\renewcommand{\baselinestretch}{1.2}\normalsize 
\centering
\begin{tabular}[t]{|l|c|c|l|c|}
\hline
			 Symmetry generator $Y'$ & $|Y' (S)|$ & $v_S = \sqrt{2}\, |\langle S\rangle|$ & Texture zeros in $\mathcal{M}_R$ & Texture zeros in $\mathcal{M}_\nu$\\
			 \hline\hline
			 $L_\mu-L_\tau$ &  $1$ & $\geq \unit[160]{GeV}$&  $(\mathcal{M}_R)_{33}$, $(\mathcal{M}_R)_{22}$ ($\vec{C}^R$) & --\\
			 $B - L_e + L_\mu - 3 L_\tau$ &  $2$ & $\geq \unit[3.5]{TeV}$& $(\mathcal{M}_R)_{33}$, $(\mathcal{M}_R)_{13}$ ($\vec{B}_4^R$)& $(\mathcal{M}_\nu)_{12}$, $(\mathcal{M}_\nu)_{22}$ ($\vec{B}_3^\nu$)\\	
			 $B - L_e - 3 L_\mu + L_\tau$ &  $2$ & $\geq \unit[4.8]{TeV}$& $(\mathcal{M}_R)_{22}$, $(\mathcal{M}_R)_{12}$ ($\vec{B}_3^R$) & $(\mathcal{M}_\nu)_{13}$, $(\mathcal{M}_\nu)_{33}$ ($\vec{B}_4^\nu$)\\
			 $B + L_e - L_\mu - 3 L_\tau$ &  $2$ & $\geq \unit[3.5]{TeV}$& $(\mathcal{M}_R)_{33}$, $(\mathcal{M}_R)_{23}$ ($\vec{D}_2^R$) & $(\mathcal{M}_\nu)_{12}$, $(\mathcal{M}_\nu)_{11}$ ($\vec{A}_1^\nu$)\\
			 $B + L_e - 3 L_\mu - L_\tau$ &  $2$ & $\geq \unit[3.5]{TeV}$& $(\mathcal{M}_R)_{22}$, $(\mathcal{M}_R)_{23}$ ($\vec{D}_1^R$) & $(\mathcal{M}_\nu)_{13}$, $(\mathcal{M}_\nu)_{11}$ ($\vec{A}_2^\nu$)\\		 
\hline
\end{tabular}
\renewcommand{\baselinestretch}{1.0}\normalsize 
\caption{
\label{tab:symmetryzeros}
Anomaly-free $U(1)$ gauge symmetries that lead to the allowed two-zero textures in the right-handed Majorana mass matrix $\mathcal{M}_R$ with the addition of just one SM singlet scalar $S$. Some of the texture zeros propagate to $\mathcal{M}_\nu \simeq -m_D \mathcal{M}_R^{-1} m_D^T$ after seesaw. Classification of the two-zero textures according to App.~\ref{sec:patterns}.}
\end{table}

Having shown that we can construct two-zero patterns via various broken flavor symmetries, we will now briefly comment on the involved scales.
The allowed two-zero textures of $\mathcal{M}_\nu^{-1}$ typically have non-vanishing entries of similar magnitude, which means that the symmetry breaking scales need to be comparable to the flavor symmetric mass terms, i.e.~$\langle S \rangle \sim \mathcal{M}_R$.
To illustrate this point, we present a particularly fascinating solution with non-vanishing elements of similar order:
\begin{align}
\mathcal{M}_R = M_0 \, \matrixx{-2 & -2 & 3\\\cdot & 1 & 0 \\ \cdot & \cdot & 0} && \Rightarrow &&
\mathcal{M}_\nu = m_0 \, \matrixx{0 & 0 & 1\\\cdot & 3 & 2 \\ \cdot & \cdot & 2} .
\end{align}
This mass matrix leads to normal hierarchy and the mixing parameters take the form
\begin{align}
\sin^2\theta_{12}\simeq \frac{1}{3}\,, &&
\sin^2 \theta_{13} = \frac{1}{3}-\frac{5}{6\sqrt{7}}\simeq 0.018\,, \\
\sin^2 \theta_{23} \simeq \frac{1}{3}+\frac{2}{3\sqrt{7}} \simeq 0.59\,,&&
\frac{\Delta m_{21}^2}{\Delta m_{31}^2} = \frac{1}{2}-\frac{5}{4\sqrt{7}} \simeq 0.027\,,
\end{align}
which are valid at $2\sigma$. It should be clear that the overall seesaw-scale is free in our models, as a change in $M_0$ can be compensated by a change in $m_D$. Thus, the predicted scaling $\langle S \rangle \sim \mathcal{M}_R$ can sit anywhere from $\unit[10^{15}]{GeV}$ to $\unit[1]{TeV}$, the latter being obviously more interesting for collider phenomenology. 

While there is no hierarchy in $\mathcal{M}_\nu$ in the case of two-zero textures, some hierarchy among the $\mathcal{M}_R$ entries is present if the elements of $m_D$, i.e.~$a$, $b$ and $c$, are hierarchical. Taking for example our model for $\vec{D}_1^R$, i.e.~$B+L_e - 3 L_\mu - L_\tau$, we find numerically the following typical hierarchy among the nonzero elements:
\begin{align}
 S_{11}/a^2 \sim S_{12}/ab \sim M_{13}/ac > S_{33}/c^2 \,.
\end{align}
Here and in the following, $M_{ij}$ denotes an $\mathcal{M}_R$ entry allowed by the imposed flavor symmetry and $S_{ij} = \lambda_{ij} \langle S \rangle$ a symmetry breaking entry. The hierarchy is very mild, but we can easily make $M_{ij} \gg S_{ij}$ by imposing $a,b\ll c$. The same qualitative result holds for $\vec{D}_2^R$. An analogous analysis of $\vec{B}_3^R$ ($B-L_e - 3 L_\mu + L_\tau$) gives
\begin{align}
 S_{11}/a^2 \gtrsim S_{23}/bc > S_{33}/c^2 \gg M_{13}/ac\,,
\end{align}
the ratio of largest to smallest nonvanishing entry being $\sim 15$. Here we can not make $M_{ij} \gg S_{ij}$, but are drawn to the scaling  $M_{ij} \sim S_{ij}$ (similar for $\vec{B}_4^R$). The same can be said for case $\vec{C}^R$ ($L_\mu-L_\tau$), with the typical relations
\begin{align}
 M_{11}/a^2 \sim S_{13}/ac \sim M_{23}/bc > S_{12}/ab\,.
\end{align}
However, for $\vec{C}^R$ there are also solutions that naturally suggest $M_{ij} > S_{ij}$.

The examples above show that the hierarchy among the $m_D$ entries reflects the hierarchy among the $\mathcal{M}_R$ entries. Similar analyses can be performed for the other patters, but the analysis will not change the conclusion that the $\mathcal{M}_R$ patterns given in Eq.~\eqref{eq:allowedtexturezerosandminors} are consistent with the most recent data within one standard deviation.

\subsection{Realization via Discrete Flavor Symmetries}

Exploiting the discrete subgroups of gauge symmetries, i.e.~discrete gauge symmetries~\cite{DGS}, is yet another way to evade the quantum gravitational breaking of discrete symmetries~\cite{DGS2}.
Hence, it may be interesting to explore discrete subgroups of the discussed $U(1)$ gauge symmetries and see whether or not they are useful to derive the allowed two-zero textures.
Taking the $B-L_e + L_\mu -3 L_\tau$ symmetry as an example, we find that its $\mathbb{Z}_5$ subgroup with a scalar with charge $3$ leads to the same phenomenology as the overlying $U(1)$. Similar discussions hold for the other symmetries that work with just one scalar. Something new happens however for the two patterns that need two scalars.
For instance, the charge matrices of the $B- L_e -5 L_\mu + 3 L_\tau$ symmetry and its $\mathbb{Z}_5$ subgroup are given by
\begin{align}
Y' (\overline{N}^c_i N_j) = \matrixx{-2 & -6 & 2\\\cdot & -10 & -2\\\cdot & \cdot & 6} ,\qquad  \matrixx{-2 & -1 & 2\\\cdot & 0 & -2\\\cdot & \cdot & 1} \bmod 5\,,
\end{align}
respectively, and we see that instead of two scalars with $|Y' (S_1)|=10$ and $|Y' (S_2)|=2$ for the $U(1)$ case, we only need one scalar with charge $2$ in the $\mathbb{Z}_5$ case to obtain the same pattern $\vec{B}_2^R$. Notice that the family non-universality is preserved even for the $\mathbb{Z}_5$ case, and thus the Dirac mass matrices remain diagonal.
In that sense, we can conclude that all allowed two-zero textures in $\mathcal{M}_R$ can be obtained from a $U(1)$ gauge symmetry, be it continuous or discrete, with just one additional complex scalar.

\subsection{Collider Opportunities}

The LHC phenomenology of the $B-\sum_\alpha x_\alpha L_\alpha$ gauge boson $Z'$ is similar to that of the $B-L$ gauge boson, so we will not discuss it here. We do however note that the LHC has the potential to differentiate between the different classes of two-zero textures in our model. The reason for the naming scheme of the two-zero textures in App.~\ref{sec:patterns} is the similar phenomenology at neutrino oscillation experiments; for example, the patterns $\vec{A}_1^\nu$ and $\vec{A}_2^\nu$ lead to almost identical predictions for the oscillation parameters and are therefore very hard to distinguish using only oscillation data.
In our framework, however, these patterns are imposed by the gauge symmetries $B + L_e - L_\mu - 3 L_\tau$ and $B + L_e - 3 L_\mu -  L_\tau$, respectively, which are much easier to separate. One just needs to look at the flavor ratios of the final state $Z'\ra \ell\ell$ to verify or exclude the different $B-\sum_\alpha x_\alpha L_\alpha$ models.

Indirect effects due to $Z$--$Z'$ mixing~\cite{Babu:1997st} can of course also be present, but are beyond the scope of this letter. The mixing of the full group $G_\sm\times \G$, e.g.~kinetic mixing of the hypercharge with $\G$ and the resulting $Z$--$Z'$--$Z''$--$Z'''$ mixing, can be discussed in analogy to Ref.~\cite{kineticmixing}.


\section{Conclusion}
\label{sec:conclusion}

The maximal anomaly-free abelian gauge group extension within the SM particle content, universal quark charges and allowed mass terms for the fermions is given by $U(1)_{L_\alpha - L_\beta}$, as determined long time ago. The introduction of three right-handed neutrinos with appropriate lepton-numbers significantly enlarges this group to
\begin{align}
 \G = U(1)_{B-L} \times U(1)_{L_e - L_\mu} \times U(1)_{L_\mu - L_\tau}\,,
\end{align}
where we choose one specific basis in group space. Lepton mixing is induced by the spontaneous breakdown of $\G$, the details of which give rise to numerous phenomenologically valid, inequivalent models. Besides the usual approximate flavor-symmetries like $L_\mu-L_\tau$ in the active neutrino mass matrix, we have shown that texture zeros and vanishing minors might also originate from $\G$.

We presented numerous examples of anomaly-free gauge symmetries that lead to two-zero textures in $\mathcal{M}_R$ and therefore to testable predictions for neutrino oscillation parameters. We showed that all allowed patterns of $\mathcal{M}_R$ ($\vec{D}_{1,2}^R$, $\vec{B}_{1,2,3,4}^R$, $\vec{C}^R$) can be implemented by a family non-universal $U(1)$ gauge symmetry with at most two new scalars, making these models very simple and possibly distinguishable at collider experiments. Using instead the discrete gauge subgroups $\mathbb{Z}_N$ of the $U(1)$ groups reduces the number of necessary new scalars to one. As a side product, we also see that four of the seven allowed two-zero textures of $\mathcal{M}_\nu$ can have this origin (including the ``least fine-tuned'' patterns $\vec{A}_i^\nu$). The remaining three two-zero textures of $\mathcal{M}_\nu$, namely $\vec{C}^\nu$, $\vec{B}_1^\nu$ and $\vec{B}_2^\nu$, can not be explained in this simple framework. 
The texture-zero approach of this letter gives rise to previously undiscussed flavor symmetries that might also be interesting in the context of baryo/leptogenesis.

\begin{acknowledgments}
The work of T.A.~was supported in part by the National Natural Science Foundation of China under grant No.~11135009 and the Chinese Academy of Sciences Fellowship for Young International Scientists.
The work of J.H.~was supported by the the ERC under the Starting Grant MANITOP and by the IMPRS-PTFS.
J.K.~is partially supported by a Grant-in-Aid for Scientific Research (C) from Japan Society for Promotion of Science (No.~2254027).
T.A.~and J.H.~thank the theory group of Kanazawa University, where most of this work was performed, for very kind hospitality.
\end{acknowledgments}

\appendix

\section{Classification of Texture Zeros and Vanishing Minors}
\label{sec:patterns}

For convenience we list the two-zero texture patterns of symmetric matrices in their common notation~\cite{moretexturezeros}:
\begin{align}
\vec{A}_1 : \quad \matrixx{0 & 0 & \times \\ 0 & \times & \times \\ \times & \times & \times} , \qquad
\vec{A}_2 : \quad \matrixx{0 & \times & 0 \\ \times & \times & \times \\ 0 & \times & \times} ;
\end{align}
\begin{align}
\vec{B}_1 : \quad \matrixx{\times & \times & 0 \\ \times & 0 & \times \\ 0 & \times & \times} , \qquad
\vec{B}_2 : \quad \matrixx{\times & 0 & \times \\ 0 & \times & \times \\ \times & \times & 0} ,\\
\vec{B}_3 : \quad \matrixx{\times & 0 & \times \\ 0 & 0 & \times \\ \times & \times & \times} , \qquad
\vec{B}_4 : \quad \matrixx{\times & \times & 0 \\ \times & \times & \times \\ 0 & \times & 0} ;
\end{align}
\begin{align}
\vec{C} : \quad \matrixx{\times & \times & \times \\ \times & 0 & \times \\ \times & \times & 0} ;
\end{align}
\begin{align}
\vec{D}_1 : \quad \matrixx{\times & \times & \times \\ \times & 0 & 0 \\ \times & 0 & \times} , \qquad
\vec{D}_2 : \quad \matrixx{\times & \times & \times \\ \times & \times & 0 \\ \times & 0 & 0} ;
\end{align}
\begin{align}
\vec{E}_1 : \quad \matrixx{0 & \times & \times \\ \times & 0 & \times \\ \times & \times & \times} , \qquad
\vec{E}_2 : \quad \matrixx{0 & \times & \times \\ \times & \times & \times \\ \times & \times & 0} , \qquad
\vec{E}_3 : \quad \matrixx{0 & \times & \times \\ \times & \times & 0 \\ \times & 0 & \times} ;
\end{align}
and
\begin{align}
\vec{F}_1 : \quad \matrixx{\times & 0 & 0 \\ 0 & \times & \times \\ 0 & \times & \times} , \qquad
\vec{F}_2 : \quad \matrixx{\times & 0 & \times \\ 0 & \times & 0 \\ \times & 0 & \times} , \qquad
\vec{F}_3 : \quad \matrixx{\times & \times & 0 \\ \times & \times & 0 \\ 0 & 0 & \times} .
\end{align}
In all cases, the symbol $\times$ denotes a non-vanishing entry. In some cases, the texture zeros propagate to the inverse matrix, we list all such patterns in Tab.~\ref{tab:patterns}. In any case, two-zero textures of $M$ lead to two vanishing minors in $M^{-1}$. We define the minor $(i,j)$ of an $n\times n$ matrix $A$ as the determinant of the $(n-1)\times (n-1)$ matrix obtained from $A$ by removing the $i$-th row and $j$-th column. This is a useful convention, because now the texture zeros $M_{ij}=0=M_{nm}$ result in the vanishing minors $(i,j)$ and $(n,m)$ of $M^{-1}$. We can therefore classify vanishing minors in $\mathcal{M}_\nu$ as texture zeros in $\mathcal{M}_\nu^{-1}$ and vice versa. Two-zero texture patterns $\vec{P}_i$ in $\mathcal{M}_\nu$ ($\mathcal{M}_R$) will be denoted with an index $\nu$ ($R$), i.e.~as $\vec{P}_i^\nu$ ($\vec{P}_i^R$), to avoid confusion between the patterns.

\begin{table}[t]
\renewcommand{\baselinestretch}{1.2}\normalsize 
\centering
\begin{tabular}[t]{|l|c|c|c|c|c|c|c|}
\hline
	Pattern of $M$ & $\vec{A}_1$& $\vec{A}_2$& $\vec{B}_3$& $\vec{B}_4$& $\vec{D}_1$& $\vec{D}_2$ & $\vec{F}_j$\\
\hline
	Pattern of $M^{-1}$  & $\vec{D}_2$& $\vec{D}_1$& $\vec{B}_4$& $\vec{B}_3$& $\vec{A}_2$& $\vec{A}_1$ & $\vec{F}_j$\\
\hline
\end{tabular}
\renewcommand{\baselinestretch}{1.0}\normalsize 
\caption{
\label{tab:patterns}
Two-texture zeros of a non-singular symmetric $3\times 3$ matrix $M$ that lead to two-texture zeros in the inverse matrix $M^{-1}$.}
\end{table}

\section{Two-Zero Textures with Two/Three Scalars}
\label{sec:lists}

In this appendix we present an exhaustive list of $U(1)$ flavor symmetries that lead to valid two-zero textures in $\mathcal{M}_R$ with the addition of two SM singlet scalars (Tab.~\ref{tab:B-xLexamples2}). All of them are of the type $B - x_e L_e - x_\mu L_\mu - (3-x_e-x_\mu)L_\tau$, because $y_e L_e + y_\mu L_\mu - (y_e + y_\mu)L_\tau$ requires at least three scalars (complete list in Tab.~\ref{tab:yLexamples2}). As can be seen in Tab.~\ref{tab:B-xLexamples2}, all patterns that already work with just one scalar (Tab.~\ref{tab:symmetryzeros}) have infinitely many realizations once another scalar is introduced. Since the patterns $\vec{D}_1^R$, $\vec{D}_2^R$, $\vec{B}_3^R$ and $\vec{B}_4^R$ are related by $L_\alpha \leftrightarrow L_\beta$ operations, we do not list them explicitly.

\begin{table}[ht]
\renewcommand{\baselinestretch}{1.2}\normalsize 
\begin{tabular}{|c|l|c|}
\hline
 $\mathcal{M}_R$ pattern & Symmetry generator $Y'$ & $|Y' (S_i)|$ \\ \hline\hline
 $\vec{D}_1^R$ & $B -a L_e - 3 L_\mu + a L_\tau$,  $a \notin \{-9,-3,0,1,3\}$ & $2|a|,~|3+a|$\\
   & $B - 2 L_\mu- L_\tau$ & $1,~2$\\
 & $B +\frac{3}{2} L_e - \frac{9}{2} L_\mu$ & $3,~\frac{3}{2}$\\
 & $B +\frac{9}{7} L_e - \frac{27}{7} L_\mu- \frac{3}{7} L_\tau$ & $\frac{18}{7},~\frac{6}{7}$\\
 & $B +\frac{1}{3} L_e - \frac{7}{3} L_\mu - L_\tau$ & $2,~\frac{2}{3}$ \\\hline
 $\vec{D}_2^R$ &  $\vec{D}_1^R$ with $L_\mu \leftrightarrow L_\tau$ & \\\hline
 $\vec{B}_3^R$ &  $\vec{D}_1^R$ with $L_e \leftrightarrow L_\tau$ & \\\hline
 $\vec{B}_4^R$ &  $\vec{B}_3^R$ with $L_\mu \leftrightarrow L_\tau$ & \\\hline
 $\vec{B}_1^R$ & $B + 3 L_\mu -6 L_\tau$ & $3,~12$\\
   & $B - 2 L_\mu- L_\tau$ & $2,~3$\\
 & $B -\frac{9}{2} L_\mu + \frac{3}{2} L_\tau$ & $3,~\frac{9}{2}$\\
 & $B -6 L_e +3 L_\mu$ & $3,~12$\\
 & $B +\frac{3}{2} L_e - \frac{9}{2} L_\mu$ & $3,~\frac{9}{2}$\\
 & $B -L_e - 2 L_\mu$ & $2,~3$\\
 & $B - L_e +3 L_\mu - 5 L_\tau$ & $2,~10$\\
 & $B -5 L_e +3 L_\mu -L_\tau$ & $2,~10$\\\hline
 $\vec{B}_2^R$ &  $\vec{B}_1^R$ with $L_\mu \leftrightarrow L_\tau$ & \\\hline
$\vec{C}^R$ & $B +3 L_e - a L_\mu -(6-a) L_\tau$,  $a \notin \{-3,0,1,3,5,6,9\}$ & $5,~|3-a|$\\
   & $B - 6 L_\mu +3 L_\tau$ & $3,~6$\\
 & $B -3 L_e \pm 9 L_\mu \mp 9 L_\tau$ & $6,~12$\\
\hline
\end{tabular}
\renewcommand{\baselinestretch}{1.0}\normalsize
\caption{$Y^\prime = B - x_e L_e - x_\mu L_\mu - (3-x_e-x_\mu)L_\tau$ charge assignments that lead to the allowed two-zero textures in the right-handed Majorana neutrino mass matrix $\mathcal{M}_R$ with two SM gauge singlet scalars. }
\label{tab:B-xLexamples2}
\end{table}

\begin{table}[ht]
\renewcommand{\baselinestretch}{1.2}\normalsize 
\begin{tabular}{|c|l|c|}
\hline
 $\mathcal{M}_R$ pattern & Symmetry generator $Y'$ & $|Y' (S_i)|$ \\ \hline\hline
 $\vec{D}_1^R$ & $L_e - 3 L_\mu + 2 L_\tau$ & $2,~3,~4$\\
   & $L_e +2 L_\mu -3 L_\tau$ & $2,~3,~6$\\
   & $3 L_e - 2 L_\mu - L_\tau$ & $1,~2,~6$\\\hline
 $\vec{D}_2^R$ &  $\vec{D}_1^R$ with $L_\mu \leftrightarrow L_\tau$ & \\\hline
 $\vec{B}_3^R$ &  $\vec{D}_1^R$ with $L_e \leftrightarrow L_\tau$ & \\\hline
 $\vec{B}_4^R$ &  $\vec{B}_3^R$ with $L_\mu \leftrightarrow L_\tau$ & \\\hline
 $\vec{B}_1^R$ & $L_e -3 L_\mu +2 L_\tau$ & $1,~2,~4$\\
   & $2 L_e -3 L_\mu + L_\tau$ & $1,~2,~4$\\\hline
 $\vec{B}_2^R$ &  $\vec{B}_1^R$ with $L_\mu \leftrightarrow L_\tau$ & \\\hline
$\vec{C}^R$ & $L_e -3 L_\mu + 2 L_\tau$ & $1,~2,~3$\\
 & $ L_e +2 L_\mu -3 L_\tau$ & $1,~2,~3$\\
\hline
\end{tabular}
\renewcommand{\baselinestretch}{1.0}\normalsize
\caption{$Y^\prime = y_e L_e + y_\mu L_\mu - (y_e + y_\mu)L_\tau$ charge assignments that lead to the allowed two-zero textures in the right-handed Majorana neutrino mass matrix $\mathcal{M}_R$ with three SM gauge singlet scalars. }
\label{tab:yLexamples2}
\end{table}

\newpage



\begin{thebibliography}{99}


\bibitem{non-uni}
  P.~Langacker and M.~Plumacher,
  Phys.\ Rev.\ D {\bf 62}, 013006 (2000)
  [\href{http://arxiv.org/abs/hep-ph/0001204}{arXiv:hep-ph/0001204}];\\
  G.~Buchalla, G.~Hiller and G.~Isidori,
  Phys.\ Rev.\ D {\bf 63}, 014015 (2000)
  [\href{http://arxiv.org/abs/hep-ph/0006136}{arXiv:hep-ph/0006136}];\\
  V.~Barger, C.~W.~Chiang, P.~Langacker and H.~S.~Lee,
  Phys.\ Lett.\ B {\bf 580}, 186 (2004)
  [\href{http://arxiv.org/abs/hep-ph/0310073}{arXiv:hep-ph/0310073}];\\
  V.~Barger, C.~W.~Chiang, P.~Langacker and H.~S.~Lee,
  Phys.\ Lett.\ B {\bf 598}, 218 (2004)
  [\href{http://arxiv.org/abs/hep-ph/0406126}{arXiv:hep-ph/0406126}];\\
  K.~Cheung, C.~W.~Chiang, N.~G.~Deshpande and J.~Jiang,
  Phys.\ Lett.\ B {\bf 652}, 285 (2007)
  [\href{http://arxiv.org/abs/hep-ph/0604223}{arXiv:hep-ph/0604223}];\\
  E.~Golowich, J.~Hewett, S.~Pakvasa and A.~A.~Petrov,
  Phys.\ Rev.\ D {\bf 76}, 095009 (2007)
  [\href{http://arxiv.org/abs/0705.3650}{arXiv:0705.3650} [hep-ph]];\\
  V.~Barger, L.~L.~Everett, J.~Jiang, P.~Langacker, T.~Liu and C.~E.~M.~Wagner,
  JHEP {\bf 0912}, 048 (2009)
  [\href{http://arxiv.org/abs/0906.3745}{arXiv:0906.3745} [hep-ph]];\\
  J.~Y.~Liu, Y.~Tang and Y.~L.~Wu,
  J.\ Phys.\ G {\bf 39}, 055003 (2012)
  [\href{http://arxiv.org/abs/1108.5012}{arXiv:1108.5012} [hep-ph]].


\bibitem{type1}
  P.~Minkowski,
  Phys.\ Lett.\  B {\bf 67}, 421 (1977);\\
  M.~Gell-Mann, P.~Ramond, and R.~Slansky, in {\it Supergravity}, p.~315, edited by P. van Nieuwenhuizen and D. Freedman, North Holland, Amsterdam, 1979;\\
  T.~Yanagida, Proc. of the {\it Workshop on Unified Theory and the Baryon Number of the Universe}, KEK, Japan, 1979;\\
  R.N. Mohapatra and G.~Senjanovi{\'c}, Phys. Rev. Lett. {\bf 44}, 912 (1980).


\bibitem{0minr}
For vanishing minors in the active neutrino mass matrix, see \\
  L.~Lavoura,
  Phys.\ Lett.\ B {\bf 609}, 317 (2005)
  [\href{http://arxiv.org/abs/hep-ph/0411232}{arXiv:hep-ph/0411232}];\\
  E.~I.~Lashin and N.~Chamoun,
  Phys.\ Rev.\ D {\bf 78}, 073002 (2008)
  [\href{http://arxiv.org/abs/0708.2423}{arXiv:0708.2423} [hep-ph]].


\bibitem{0minr2}
  S.~Dev, S.~Gupta, R.~R.~Gautam and L.~Singh,
  Phys.\ Lett.\ B {\bf 706}, 168 (2011)
  [\href{http://arxiv.org/abs/1111.1300}{arXiv:1111.1300} [hep-ph]].


\bibitem{moretexturezeros}
For texture zeros in the active neutrino mass matrix, see \\
  M.~S.~Berger and K.~Siyeon,
  Phys.\ Rev.\ D {\bf 64}, 053006 (2001)
  [\href{http://arxiv.org/abs/hep-ph/0005249}{arXiv:hep-ph/0005249}];\\
  P.~H.~Frampton, S.~L.~Glashow and D.~Marfatia,
  Phys.\ Lett.\ B {\bf 536}, 79 (2002)
  [\href{http://arxiv.org/abs/hep-ph/0201008}{arXiv:hep-ph/0201008}];\\
  P.~H.~Frampton, M.~C.~Oh and T.~Yoshikawa,
  Phys.\ Rev.\ D {\bf 66}, 033007 (2002)
  [\href{http://arxiv.org/abs/hep-ph/0204273}{arXiv:hep-ph/0204273}];\\
  A.~Kageyama, S.~Kaneko, N.~Shimoyama and M.~Tanimoto,
  Phys.\ Lett.\ B {\bf 538}, 96 (2002)
  [\href{http://arxiv.org/abs/hep-ph/0204291}{arXiv:hep-ph/0204291}];\\
  E.~Ma,
  Phys.\ Rev.\ D {\bf 71}, 111301 (2005)
  [\href{http://arxiv.org/abs/hep-ph/0501056}{arXiv:hep-ph/0501056}];\\
  C.~Hagedorn and W.~Rodejohann,
  JHEP {\bf 0507}, 034 (2005)
  [\href{http://arxiv.org/abs/hep-ph/0503143}{arXiv:hep-ph/0503143}];\\
  N.~Haba and K.~Yoshioka,
  Nucl.\ Phys.\ B {\bf 739}, 254 (2006)
  [\href{http://arxiv.org/abs/hep-ph/0511108}{arXiv:hep-ph/0511108}];\\
  A.~Dighe and N.~Sahu,
  \href{http://arxiv.org/abs/0812.0695}{arXiv:0812.0695} [hep-ph];\\
  S.~Dev, S.~Verma, S.~Gupta and R.~R.~Gautam,
  Phys.\ Rev.\ D {\bf 81}, 053010 (2010)
  [\href{http://arxiv.org/abs/1003.1006}{arXiv:1003.1006} [hep-ph]];\\
  S.~Verma,
  Nucl.\ Phys.\ B {\bf 854}, 340 (2012)
  [\href{http://arxiv.org/abs/1109.4228}{arXiv:1109.4228} [hep-ph]];\\
  P.~M.~Ferreira and L.~Lavoura,
  \href{http://arxiv.org/abs/1202.4024}{arXiv:1202.4024} [hep-ph].


\bibitem{0texr}
For a recent analysis of texture zeros in the active neutrino mass matrix, see \\
  H.~Fritzsch, Z.~Z.~Xing and S.~Zhou,
  JHEP {\bf 1109}, 083 (2011)
  [\href{http://arxiv.org/abs/1108.4534}{arXiv:1108.4534} [hep-ph]].


\bibitem{Zn}
  W.~Grimus, A.~S.~Joshipura, L.~Lavoura and M.~Tanimoto,
  Eur.\ Phys.\ J.\ C {\bf 36}, 227 (2004)
  [\href{http://arxiv.org/abs/hep-ph/0405016}{arXiv:hep-ph/0405016}];\\
  S.~Dev, S.~Gupta and R.~R.~Gautam,
  Phys.\ Lett.\ B {\bf 701}, 605 (2011)
  [\href{http://arxiv.org/abs/1106.3451}{arXiv:1106.3451} [hep-ph]].


\bibitem{gravity}
  S.~W.~Hawking,
  Phys.\ Lett.\ B {\bf 195}, 337 (1987);\\
  G.~V.~Lavrelashvili, V.~A.~Rubakov and P.~G.~Tinyakov,
  JETP Lett.\  {\bf 46}, 167 (1987);\\
  S.~B.~Giddings and A.~Strominger,
  Nucl.\ Phys.\ B {\bf 306}, 890 (1988);\\
  S.~R.~Coleman,
  Nucl.\ Phys.\ B {\bf 310}, 643 (1988).


\bibitem{dwall}
  T.~W.~B.~Kibble,
  J.\ Phys.\ A {\bf 9}, 1387 (1976).


\bibitem{B-stuff}
  E.~Ma,
  Phys.\ Lett.\  B {\bf 433}, 74 (1998)
  [\href{http://arxiv.org/abs/hep-ph/9709474}{arXiv:hep-ph/9709474}];\\
  E.~Ma and D.~P.~Roy,
  Phys.\ Rev.\  D {\bf 58}, 095005 (1998)
  [\href{http://arxiv.org/abs/hep-ph/9806210}{arXiv:hep-ph/9806210}];\\
  E.~Ma and U.~Sarkar,
  Phys.\ Lett.\  B {\bf 439}, 95 (1998)
  [\href{http://arxiv.org/abs/hep-ph/9807307}{arXiv:hep-ph/9807307}];\\
  E.~Ma, D.~P.~Roy and U.~Sarkar,
  Phys.\ Lett.\  B {\bf 444}, 391 (1998)
  [\href{http://arxiv.org/abs/hep-ph/9810309}{arXiv:hep-ph/9810309}];\\
  E.~Ma and D.~P.~Roy,
  Phys.\ Rev.\  D {\bf 59}, 097702 (1999)
  [\href{http://arxiv.org/abs/hep-ph/9811266}{arXiv:hep-ph/9811266}];\\
  E.~Salvioni, A.~Strumia, G.~Villadoro and F.~Zwirner,
  JHEP {\bf 1003}, 010 (2010)
  [\href{http://arxiv.org/abs/0911.1450}{arXiv:0911.1450} [hep-ph]];\\
  H.~S.~Lee and E.~Ma,
  Phys.\ Lett.\ B {\bf 688}, 319 (2010)
  [\href{http://arxiv.org/abs/1001.0768}{arXiv:1001.0768} [hep-ph]].


\bibitem{Davoudiasl:2011sz} 
  H.~Davoudiasl, H.~S.~Lee and W.~J.~Marciano,
  Phys.\ Rev.\ D {\bf 84}, 013009 (2011)
  [\href{http://arxiv.org/abs/1102.5352}{arXiv:1102.5352} [hep-ph]].


\bibitem{zero}
  R.~Foot,
  Mod.\ Phys.\ Lett.\  A {\bf 6}, 527 (1991);\\
  X.~G.~He, G.~C.~Joshi, H.~Lew and R.~R.~Volkas,
  Phys.\ Rev.\  D {\bf 44}, 2118 (1991);\\
  R.~Foot, X.~G.~He, H.~Lew and R.~R.~Volkas,
  Phys.\ Rev.\  D {\bf 50}, 4571 (1994)
  [\href{http://arxiv.org/abs/hep-ph/9401250}{arXiv:hep-ph/9401250}].


\bibitem{Lmu-Ltau}
See for example
  J.~Heeck and W.~Rodejohann,
  Phys.\ Rev.\ D {\bf 84}, 075007 (2011)
  [\href{http://arxiv.org/abs/1107.5238}{arXiv:1107.5238} [hep-ph]].


\bibitem{kineticmixing} 
  J.~Heeck and W.~Rodejohann,
  Phys.\ Lett.\ B {\bf 705}, 369 (2011)
  [\href{http://arxiv.org/abs/1109.1508}{arXiv:1109.1508} [hep-ph]].


\bibitem{Choubey:2004hn} 
  S.~Choubey and W.~Rodejohann,
  Eur.\ Phys.\ J.\ C {\bf 40}, 259 (2005)
  [\href{http://arxiv.org/abs/hep-ph/0411190}{arXiv:hep-ph/0411190}].


\bibitem{soon}
  J.~Heeck and W.~Rodejohann,
  Phys.\  Rev.\ D {\bf 85}, 113017 (2012)
  [\href{http://arxiv.org/abs/1203.3117}{arXiv:1203.3117} [hep-ph]].


\bibitem{Froggatt:1978nt} 
  C.~D.~Froggatt and H.~B.~Nielsen,
  Nucl.\ Phys.\ B {\bf 147}, 277 (1979).


\bibitem{globalfit}
  T.~Schwetz, M.~Tortola and J.~W.~F.~Valle,
  New J.\ Phys.\  {\bf 13}, 063004 (2011)
  [\href{http://arxiv.org/abs/1103.0734}{arXiv:1103.0734} [hep-ph]];\\
  T.~Schwetz, M.~Tortola and J.~W.~F.~Valle,
  New J.\ Phys.\  {\bf 13}, 109401 (2011)
  [\href{http://arxiv.org/abs/1108.1376}{arXiv:1108.1376} [hep-ph]].


\bibitem{dayabay} 
  F.~P.~An {\it et al.} [Daya Bay Collaboration],
  Phys.\ Rev.\ Lett.\  {\bf 108}, 171803 (2012)
  [\href{http://arxiv.org/abs/1203.1669}{arXiv:1203.1669} [hep-ex]].


\bibitem{finetuning} 
  D.~Meloni and G.~Blankenburg,
  \href{http://arxiv.org/abs/1204.2706}{arXiv:1204.2706} [hep-ph].


\bibitem{one_vanishing_minor} 
  E.~I.~Lashin and N.~Chamoun,
  Phys.\ Rev.\ D {\bf 80}, 093004 (2009)
  [\href{http://arxiv.org/abs/0909.2669}{arXiv:0909.2669} [hep-ph]].


\bibitem{LEP-2bounds}
  [The LEP Collaborations: ALEPH Collaboration, DELPHI Collaboration, L3 Collaboration,
OPAL Collaboration, the LEP Electroweak Working Group, the SLD Electroweak, Heavy
Flavour Groups],
  \href{http://arxiv.org/abs/hep-ex/0312023}{arXiv:hep-ex/0312023};\\
  M.~S.~Carena, A.~Daleo, B.~A.~Dobrescu and T.~M.~P.~Tait,
  Phys.\ Rev.\  D {\bf 70}, 093009 (2004)
  [\href{http://arxiv.org/abs/hep-ph/0408098}{arXiv:hep-ph/0408098}].


\bibitem{DGS}
  F.~J.~Wegner, 
  J.\ Math.\ Phys. {\bf 12}, 2259 (1971).


\bibitem{DGS2}
  L.~M.~Krauss and F.~Wilczek, 
  Phys.\ Rev.\ Lett. {\bf 62}, 1221 (1989).


\bibitem{Babu:1997st} 
  K.~S.~Babu, C.~F.~Kolda and J.~March-Russell,
  Phys.\ Rev.\ D {\bf 57}, 6788 (1998)
  [\href{http://arxiv.org/abs/hep-ph/9710441}{arXiv:hep-ph/9710441}].


\end{thebibliography}
\end{document}